\documentclass{nature}

\title{The Next U.S. Astronomy Decadal Survey}

\author{Adam Burrows$^{1}$}

\begin{document}

\maketitle

\begin{affiliations}
 \item Department of Astrophysical Sciences, Princeton University, Princeton, NJ 08544
\end{affiliations}

\begin{abstract}
The U.S. astronomy decadal surveys have  
been models for advice to government on how
to apportion resources to optimise the scientific return
on national investments in facilities and manpower. The
U.S. is now gearing up to conduct its 2020 survey and the
results are likely to guide international astronomy far into the future.
Here, I summarize the current strains in an otherwise
world-leading program of ground- and space-based astronomical
discovery and some of the issues that will be faced by the 
participants in this upcoming collective exercise.
\end{abstract}

\section*{}

For more than fifty years, the U.S. astronomy community has conducted
surveys of its field every decade to determine its scientific priorities
and to help guide the federal government as it allocates resources for astronomical
research. By making the hard choices based on scientific merit and rallying the community
behind common goals, the surveys have been the gold standard for scientific 
advice to Congress and the Administration.  The results to date have  
been spectacular, leading to the construction of the original national 
observing facilities in Arizona and Chile, the Great Observatories in space 
(Hubble, Spitzer, Chandra, and the Gamma-Ray Observatory [GRO]), and the Very Large Array 
in Socorro (think ``Contact"), to name only a very few. These and other strategic investments 
enabled the explosive growth of knowledge about the Universe that continues to this day
and the discoveries that are some of the defining facts of modern international 
civilisation. 

In the study of the foamy structure of the Universe and the cosmic web of galaxies, 
we have entered an era of precision cosmology that is yielding fundamental
insights into the character of its evolution at the percent level. We have 
discovered that the Universe is dominated by dark matter ($\sim$25\%)
and dark energy ($\sim$70\%) (but do not know what either is), that matter 
as we know it constitutes but $\sim$5\%, and that the expansion of the Universe 
is in fact accelerating. Detailed measurements of the cosmic microwave 
background (from space using WMAP and Planck, and from the ground in Chile 
and the south pole) and large-scale galaxy surveys (pioneered by the Sloan Digital
Sky Survey [SDSS] and its successor efforts, and now leading to the Dark Energy Survey [DES] and
the Dark Energy Spectroscopic Instrument [DESI]) are mapping, or soon will map, the 
galactic superstructure, cosmic acceleration, and structure growth in unprecedented detail
and depth. 

The study of planets beyond the solar system (``exoplanets") has exploded with a constant flood
of discoveries by transit, radial-velocity stellar wobble, direct imaging, and microlensing
techniques, both from space (the Kepler satellite, in particular) and the ground.  
The thousands of exoplanets now known constitute a heterogeneous collection that spans
masses from sub-terrestrial to super-Jovian and orbital separations from their primaries
from the torrid climes in extremis to the cold remotenesses far beyond the warming effect
of their parent star.  Many multiple planet systems with exotic dynamical architectures 
quite unlike our solar system litter the galaxy and challenge every orthodoxy 
concerning their origins and evolution.  We now know that on average more than one planet 
orbits every star and a rich set of questions has emerged, that in the 
process has revived and energized stellar astronomy itself. 

Furthermore, recent decades have seen the introduction of very-large-format CCDs, 
powerful infrared and optical multi-object spectrographs, and X-ray gratings and 
bolometers; the maturation of adaptive optics to cancel the blurring effects of 
the atmosphere; and the integration into radio telescopes of sophisticated
correlators and Moore's-Law-enabled digital signal processing. In addition to
these technical advances are the advent of gravitational wave astronomy, 
attested to by the stunning detection by the Laser Interferomter Gravitational-Wave 
Observatory (LIGO) of the merger of two massive black holes and the demonstration of
a corresponding capability in space by the Laser Interferometer Space Antenna 
(LISA) pathfinder. The Atacama Large Millimeter/submillimeter Array 
(ALMA) has recently come online and represents 
a quantum leap in our ability to image and dissect star-forming regions and 
proto-planetary disks. Moreover, the maturation of time-domain astronomy, anticipated 
by the last astronomy decadal survey \cite{nwnh} (NWNH 2010 $-$ ``New Worlds, New 
Horizons"), has revealed an exotic transient universe, enabled the 
near-continuous gazing at large swathes of the sky in the 
optical, and presages the Large Synoptic Survey Telescope's (LSST's) both 
deep and wide monitoring of a respectable chunk of the Universe
using a 3.2-gigapixel camera mated to an 8-meter-class primary 
mirror.   

These and other developments mark ours a singular age of astronomical 
discovery. The exponential growth of knowledge of the Universe and its 
contents that characterizes the last fifty years is a signature accomplishment 
that has enriched international culture and profoundly deepened mankind's 
understanding of its place in the Cosmos.

The U.S. astronomical community is about to engage in its next decadal survey
(due in 2020) and the prospects for continuing with its past vigor 
the astronomical revolution just described are at best uncertain.  
The budget of the Division of Astronomical Sciences of the U.S. National 
Science Foundation (NSF) is under severe strain. Having succeeded in
procuring funds for ALMA, the LSST, and the Daniel K. Inouye Solar 
Telescope (DKIST) under the Major Research Equipment and Facilities 
Construction (MREFC) program of the NSF, the Division is now saddled with 
the operations costs for these state-of-the-art facilities at a time when 
its overall budget has declined in real terms over the last decade.  
On the recommendation of NWNH, the Mid-Scale Innovations Program (MSIP) 
was established to bridge the funding gap between the small Major 
Research Instrumentation (MRI) program and the MREFC, but it has
never been fully funded. While the European ``Extremely-Large Telescope" 
(EELT) project seems on track, the two private U.S. ``thirty-meter" telescope
initiatives are struggling without federal dollars, and funding for 
ground-based instrumentation on already existing telescopes within
the perview of the NSF is wholly inadequate. Within an effectively 
shrinking budget, the growth in its facilities and operations costs
and the growth in the number of submitted proposals are putting the crucial 
individual investigator grants program at risk. The smallest proposal 
success rate within memory is only one consequence. To have a viable future,
the NSF must be able to divest itself expeditiously of old facilities,
preferably by selling them to interested private parties, obtain sufficient 
federal funds to operate its new facilities, and be put on a funding 
trajectory that ensures it can fulfill its central role as a major catalyst
for astronomical discovery\cite{midterm}.

 The Astrophysics Division of NASA's Science Mission Directorate is 
responsible for space telescopes, both large and small, that 
probe the cosmos above our otherwise obscuring atmosphere. Its
budget outlook is not as dire as that of the NSF Astronomy,
but there are numerous storm clouds on its horizon. The Hubble 
Space Telescope (HST; 2.4-meter aperture) is one of its iconic platforms 
and it is soon to be followed by the James Webb Space Telescope 
(JWST; 6.5-meter aperture). JWST will be much more capable than even HST,
but its costs ballooned until they were brought under control a few years ago.
Nevertheless, the result was a severe belt tightening in the Astrophysics Division that
mortgaged its near-term future and compromised the execution of NWNH.  
In particular, X-ray and gravitational wave missions that were highly 
ranked in NWNH were ceded to Europe, and thereby reduced in scope.
The esteemed Explorer program, with its mandate of a high cadence of 
smaller missions, was scaled back from what was recommended by
the 2010 Decadal Survey. WFIRST, the one flagship mission to have 
emerged from NWNH, was not envisioned a flagship mission. It will 
be a deep and wide optical/infrared survey instrument that will map 
in unprecedented detail the large-scale structure of the Universe 
to extract fundamental cosmological parameters.  It is also designed 
to discover microlensing events due to exoplanets and obtain images
of exoplanets in reflection using a state-of-the-art coronagraph. 
However, its cost growth is currently under active review and scrutiny,
as NASA tries to avoid yet another ``JWST trap" that sacrifices 
the small and medium initiatives for the large. 

Given these stresses on U.S. astronomy and astrophysics budgets, much
is riding on the upcoming survey. What are the lessons learned \cite{lessons}? 
How can the 2020 survey provide optimal advice and guidance?
Foremost, it must articulate an exciting scientific vision.
Next, it must fully engage the astronomy 
community from the outset and during the process. If astronomers have 
unrealistic expectations, are unaware of the real budget constraints,
and do not support the product, they will be problematic allies.
Astronomy decadal surveys in the past have resonated with Congress precisely
because they did the hard work of vetting alternatives in advance and of designing 
a workable and inspiring scientific strategy around which the community could rally. 
Furthermore, it will be more crucial this round than in the past to cost
proposed initiatives and missions. The CATE (Cost And Technical Evaluations) 
process of 2010 was a new departure, and it will be a central
feature of 2020.  However, while costing space-based initiatives
is a developed art, performing the same exercise for ground-based
facilities is not. How to do this credibly and usefully will be an 
issue. In addition, a CATE process with feedback from the cognizant 
technical community and from those proposing a mission
would make the costing and evaluation enterprise more robust.  
Including estimates of full life-cycle costs will be essential if
we are to avoid the damaging squeeze of operations cost growth. 

A decadal survey is nothing if not an exercise in prioritisations,
but when budgets are not up to expectations decision points and off-ramps, 
crafted by an informed committee, can make emerging budget exigencies
less painful and more science-based.  These will need to be addressed
deftly and with finesse in 2020.  There are differing opinions
about whether the branch points should be solely science-based,
or should include budget allocation priorities in the event of budget 
shortfalls. This is all the more germane when striving to maintain
balance in the overall program between small, medium, and large
initiatives and missions. The tendency of large missions (such as JWST
and, now, WFIRST) to expand at the expense of the rest of the portfolio
is all too real, and has led in the past to distortions that should be
resisted.  

Finally, as the space- and ground-based programs of Europe, Japan, 
Canada, India, China, and Russia have matured, the need to ensure 
the U.S. Decadal Survey is internationally relevant is an even more palpable reality. 
Most large missions and facilities (such as JWST, ALMA, LISA, LIGO/Virgo, to 
name but a few) have an international character and dynamic, and
a significant portion of forefront astronomical research is conducted outside 
the U.S. The 2020 survey will do an important service if it can creatively address the 
issues surrounding the coordination of the different international programs and 
collaborations, with their unique financial and political contexts.
The charge of 2020 will be no less than to optimise the science return of 
the significant national investments being made with an exciting and realisable 
agenda of astronomical discovery. If the past is any guide, the science 
case should be the easy part.

\begin{addendum}
 \item The author declares that he has no
competing financial interests. Correspondence and requests for materials
should be addressed to Adam Burrows~(email: burrows@astro.princeton.edu).
\end{addendum}


\begin{thebibliography}{1}
\bibitem{nwnh} National Research Council (NRC), 2010, ``New Worlds, New 
Horizons in Astronomy and Astrophysics", The National Academies Press, Washington, D.C.
\bibitem{midterm} National Academies of Sciences, Engineering, and Medicine, 2016.
``New Worlds, New Horizons: A Midterm Assessment". Washington, D.C.: 
The National Academies Press. doi:10.17226/23560.
%Committee on the Review of Progress Toward the Decadal Survey Vision in New Worlds, 
%New Horizons in Astronomy and Astrophysics, The National Academies Press, Washington, D.C.
\bibitem{lessons} National Academies of Sciences, Engineering, and Medicine, 2015,
``The Space Sciecne Decadal Surveys: Lessons Learned and Best Practices", 
The National Academies Press, Washington, D.C.
\end{thebibliography}
\end{document}